\newcommand{\no}{\nonumber\\}
\newcommand{\pd}{\partial}
\def\({\left(}
\def\){\right)}
\def\<{\langle}
\def\>{\rangle}

\def\be{\begin{equation}}
\def\ee{\end{equation}}
\def\bea{\begin{eqnarray}}
\def\eea{\end{eqnarray}}

\def\k{{\kappa}}
\def\l{{\lambda}}

\def\d{{\delta}}

\def\o{{\omega}}
\def\O{{\Omega}}
\def\e{{\epsilon}}

\def\a{{\alpha}}
\def\b{{\beta}}

\def\g{{\gamma}}
\def\G{{\Gamma}}
\def\p{{\pi}}
\def\P{{\Pi}}
\def\m{{\mu}}
\def\n{{\nu}}
\def\r{{\rho}}
\def\s{{\sigma}}

\def\ps{{\psi}}

\def\x{{\xi}}
\def\P{{\Pi}}
\def\0{{(0)}}
\def\1{{(1)}}
\def\2{{(2)}}

\def\dg{\dagger}

\def\hb{\hbar}

\def\mJ{{\mathcal{J}}}
\def\mV{{\mathcal{V}}}
\def\mA{{\mathcal{A}}}
\def\mP{{\mathcal{P}}}
\def\mF{{\mathcal{F}}}
\def\mS{{\mathcal{S}}}
\def\mK{{\mathcal{K}}}
\def\mD{{\mathcal{D}}}

\documentclass[preprint,unsortedaddress,amsmath,amssymb]{revtex4-1}
\usepackage{bm}
\usepackage{graphicx,color}
\usepackage{slashed}
\usepackage{ulem}
\usepackage{mathrsfs}

\date{\today}

\begin{document}

\title{\bf Quantum Kinetic Theory with Vector and Axial Gauge Fields}

\author{Zhou Chen}
\email{chenzh339@mail2.sysu.edu.cn}
\author{Shu Lin}
\email{linshu8@mail.sysu.edu.cn}
\affiliation{School of Physics and Astronomy, Sun Yat-Sen University, Zhuhai 519082, China}

\begin{abstract}
In this paper we introduce the axial gauge field to the framework of the quantum kinetic theory with vector gauge field in the massless limit. Treating axial-gauge field on an equal footing with the vector-gauge field, we construct a consistent solution to the kinetic equations up to the first order in gradient expansion or equivalently the semi-classical expansion. The intuitive extension of quantum kinetic theory presented in this work provides a natural generalization, and turns out to give rise to the covariant anomaly and the covariant currents. The corresponding consistent currents can be obtained from the covariant ones by adding the Chern-Simons current. We use the consistent currents to calculate various correlation functions among currents and energy-momentum tensor in equilibrium state.
\end{abstract}

\maketitle


\newpage

\section{Introduction}

The quantum kinetic theories for spinning particles has received much attention over the past few years. The most prominent one is quantum kinetic theory for spin one half particle, which has been widely used in the studies of spin sensitive transports. Its massless limit, chiral kinetic theory(CKT) \cite{Son:2012zy,Stephanov:2012ki,Pu:2010as,Chen:2012ca,Hidaka:2016yjf,Manuel:2013zaa,Manuel:2014dza,Huang:2018wdl,Carignano:2018gqt,Gao:2018wmr,Wang:2019moi,Lin:2019ytz,Gao:2019zhk,Hayata:2020sqz,Yang:2020mtz,Lin:2019fqo,Lin:2021sjw,Luo:2021uog}, has provided a novel description of the celebrated chiral magnetic effect \cite{Kharzeev:2004ey,Kharzeev:2007tn,Fukushima:2008xe} and chiral vortical effect \cite{Erdmenger:2008rm,Banerjee:2008th,Neiman:2010zi,Landsteiner:2011cp}. More recently, generalization to the massive case, axial kinetic theory(AKT) \cite{Hattori:2019ahi,Weickgenannt:2019dks,Gao:2019znl,Liu:2020flb,Guo:2020zpa}, has revealed additional degrees of freedom not present in the CKT. The inclusion of mass is a crucial step towards realistic description of particle polarization. The axial kinetic theory has been applied to physics of spin polarization in heavy ion collisions, see \cite{Gao:2020vbh} for a review. Collisional effects have been studied in \cite{Hidaka:2016yjf,Zhang:2019xya,Li:2019qkf,Carignano:2019zsh,Yang:2020hri,Wang:2020pej,Shi:2020htn,Weickgenannt:2020aaf,Shi:2020htn,Hou:2020mqp,Yamamoto:2020zrs,Weickgenannt:2021cuo,Sheng:2021kfc,Wang:2021qnt,Lin:2021mvw}. Quantum kinetic theory for particles with other spins have been also constructed \cite{Huang:2018aly,Huang:2020kik,Hattori:2020gqh}.

The quantum kinetic theory also offers a way to derive hydrodynamics from a microscopic theory. A key element of hydrodynamics is the response of spin one half particles to external sources, including vector gauge field and torsionful metric. These allow for derivation of anomalous hydrodynamics \cite{Satow:2014lia,Gorbar:2017toh,Yang:2020mtz,Buzzegoli:2017cqy,Buzzegoli:2018wpy}, magnetohydrodynamics \cite{Lin:2021sjw} and spin hydrodynamics \cite{Florkowski:2018fap,Bhadury:2020puc,Shi:2020htn,Peng:2021ago}\footnote{The derivation based on responses works for more general systems where kinetic description might not apply, see \cite{Banerjee:2012iz,Hernandez:2017mch,Gallegos:2021bzp,Hongo:2021ona} and references therein}. In fact, a more complete anomalous hydrodynamics also include response to axial gauge field \cite{Son:2009tf,Neiman:2010zi}.
It is desirable to extend the present framework of quantum kinetic theory to incorporate axial gauge field as well. The purpose of introducing axial gauge field is twofold. On one hand, while axial gauge field is not a physical gauge field, it can be mimicked in system like Weyl semi-metal leading to physical effects \cite{Zyuzin:2012vn}. On the other hand, the axial gauge field can also serve as a convenient tool for deriving correlation functions of axial current. In the same spirit, vierbein and spin connection in torsionful metric allows for derivation of correlation functions of energy-momentum tensor and spin tensor \cite{Hongo:2021ona}. Since spin tensor and axial current are related, our approach can be an alternative to theory with torsionful metric.

As is well known, introducing the axial gauge field leads to ambiguity in the definition of current: consistent current versus covariant current, see \cite{Landsteiner:2016led} for a review.  The purpose of this study is to integrate axial field in the framework of quantum kinetic theory. As we shall see, it is convenient to treat vector/axial gauge fields on the equal footing. This corresponds to working with covariant current. The derivation of correlation functions uses consistent current instead. We will illustrate how to calculate correlation functions with simple examples.

This paper is structured as follows: in Section \ref{sec_2}, we derive chiral kinetic theory with vector/axial gauge field in the collisionless limit; in Section \ref{sec_3}, we present solution of the kinetic equation up to first order in gradient; Section \ref{sec_4} is devoted to the calculation of one-point functions, which give rise to anomalous transports and multi-point correlation functions. We summarize and provide an outlook in Section \ref{sec_5}. Detail of calculations is left to two appendices.

\section{Quantum Kinetic Equations with Vector/Axial Gauge Fields}\label{sec_2}
With both the vector and axial gauge fields, the chiral fermion Lagrangian with the gauge fields as backgrounds is given by
\be\label{b1}
 \mathcal{L}=i\bar\psi\slashed{D}\ps.
  \ee
$D_\mu=\pd_\mu+iA_\mu+i \g^5 A^5_\mu$ is the extended covariant derivative, with $A_\mu$ and $A^5_\mu$ being the vector and axial gauge potential, respectively. And the coupling constants have been absorbed into the corresponding gauge potentials. In the current work, we restrict ourselves to the collisionless limit.
The kinetic equation is formulated in terms of the Wigner function defined as
\begin{align}
\overline{S}_{\a\b}^<(x,y)&=\langle\bar{\psi}_\b(y)\psi_\a(x)\rangle, \no
S_{\a\b}^<(p,X)&=\int d^4s e^{ip\cdot s}\overline{S}^<_{\a\b}(x,y),
\end{align}
with $X=\frac{x+y}{2}$ and $s=x-y$. $\overline{S}_{\a\b}^<(x,y)$ satisfies the Dirac equation
\begin{align}\label{dirac}
D_x\overline{S}_{\a\b}^<(x,y)=0.
\end{align}
Assuming constant field strength for the vector and axial fields, we can Fourier transform Eq.\eqref{dirac} to obtain the EOM of $S_{\a\b}^<(p,X)$
\be\label{b2}
 \g_\m\( \mK^\m+\frac{i}{2} \mD^\m +\g^5 \Pi^\m\)S^<(p,X)=0,
  \ee
%
where $\mK_\mu=p_\mu- A_\mu$, $\mD_\mu=\pd_{X,\mu} + (\pd_{X,\n} A_\mu)\pd^\n_p$ and $\Pi_\mu=A^5_\mu +\frac{i}{2}(\pd_{X,\n}A^5_\mu)\pd^\n_p$. $\mD_\m$ and $\P_\m$ are organized as a gradient expansion in $\pd_X$\footnote{In the absence of collision term, the expansion in $\pd_X$ is equivalent to expansion in $\hb$.}, with $A_\m,A_\m^5\sim O(\pd_X^0)$. To avoid the subtleties of the axial-gauge symmetry, we proceed by solving Eq.(\ref{b2}) directly without defining any gauge link. It turns out that the gauge-linked Wigner function can be obtained from the bare one by replacing the canonical momentum $p_\mu$ by the kinetic momentum $k_\m=p_\m -A_\m\mp A_\m^5$ corresponding to right and left-handed components respectively. 
It follows that the resulting solution is invariant under both vector and axial gauge symmetry, leading to covariant vector and axial currents.

The Wigner function $S^<(p,X)$, satisfying the ``hermitian'' condition $S^<(p,X)= \g^0 {S^<(p,X)}^\dagger\g^0$, can be decomposed in terms of 16 independent generators of the Clifford algebra,
\be\label{b4}
 S^<(p,X)=\frac{1}{4}\left[\mathcal{F}+ i\gamma^5\mathcal{P}+\gamma ^\mu\mathcal{V}_\mu+ \gamma^5 \gamma^\mu\mathcal{A}_\mu +\frac{\sigma^{\mu\nu}}{2}\mathcal{S}_{\mu\nu}\right],
   \ee
with the real coefficients $\mF$, $\mP$, $\mV_\m$, $\mA_\m$ and $\mS_{\m\n}$ being its scalar, pseudo-scalar, vector, axial-vector and tensor components respectively. To lighten the notation, the phase-space coordinates $(p,X)$ of the Clifford-algebra coefficients have been omitted. We can further derive
\be\label{b5}
 \begin{split}
  \g^\m S^<(p,X)=&\frac{1}{4}\left[\mV^\m-\g^5 \mA^\m+\g_\n\(g^{\m\n}\mF+i\mS^{\m\n}\)\right.\\
                  &\left.+\g^5\g_\n\(\tilde\mS^{\m\n}-ig^{\m\n}\mP\) -\frac{\s_{\a\b}}{2}\(\e^{\m\n\a\b}\mA_\n+i\(g^{\m\a}\mV^\b-g^{\m\b}\mV^\a\)\)\right],
     \end{split}
      \ee
\be\label{b6}
 \begin{split}
  \g^5\g^\m S^<(p,X)=&\frac{1}{4}\left[-\mA^\m+\g^5 \mV^\m+\g_\n\(\tilde\mS^{\m\n}-i g^{\m\n}\mP\)\right.\\
                      &\left.+\g^5\g_\n \(g^{\m\n}\mF+i\mS^{\m\n}\) +\frac{\s_{\a\b}}{2}\(\e^{\m\n\a\b}\mV_\n+i\(g^{\m\a}\mA^\b-g^{\m\b}\mA^\a\)\)\right],
     \end{split}
      \ee
where we have used the dual tensor $\tilde\mS^{\m\n}=\frac{1}{2}\e^{\m\n\r\s}\mS_{\r\s}$ and the following identities:
\be
\g^\m\g^\a\g^\b=g^{\m\a}\g^\b-g^{\m\b}\g^\a+g^{\a\b}\g^\m-i\e^{\m\a\b\l}\g^5\g_\l,\;\;\g^5\s_{\m\n}=\frac{i}{2} \e_{\m\n\k\l}\s^{\k\l}.
\ee
Inserting Eqs.(\ref{b5}) and (\ref{b6}) into Eq.(\ref{b2}), and comparing the real and imaginary parts of the coefficients in the Clifford-algebra basis, we will obtain two sets of equations. We see that the equations for $\mA_\m$ and $\mV_\m$ are decoupled from other components,
\be
 \begin{split}
  \mK_\m\mV^\mu-A^5_\mu \mA^\mu=0,\\
   \mD_\m \mA^\mu+(\pd_{X,\l} A^5_\mu) \pd_p^\l \mV^\mu=0,\\
    \mK_\m \mA_\n-\mK_\n \mA_\m -A_\m^5\mV_\n+A^5_\n\mV_\m
    +\frac{1}{2}\e_{\m\n\a\b}\mD^\a \mV^\b +\frac{1}{2}\e_{\m\n\a\b}(\pd_\l A_5^\a)\pd_p^\l\mA^\b=0,
      \end{split}
       \ee
\bea
 \begin{split}
  \mK_\m\mA^\mu-A^5_\mu \mV^\mu=0,\\
   \mD_\m \mV^\mu+(\pd_{X,\l} A^5_\mu) \pd_p^\l \mA^\mu=0,\\
    \mK_\m \mV_\n-\mK_\n \mV_\m -A_\m^5\mA_\n+A^5_\n\mA_\m+\frac{1}{2}\e_{\m\n\a\b}\mD^\a \mA^\b +\frac{1}{2}\e_{\m\n\a\b}(\pd_\l A_5^\a)\pd_p^\l\mV^\b=0.
      \end{split}
       \eea\\
The equations can be further decoupled in the chiral basis
\be\label{b7}
 k^s_{\m}\mJ^\m_s=0,
  \ee
\be\label{b8}
 \nabla^s_{\m}\mJ^\m_s=0,
  \ee
\be\label{b9}
 k^\m_s\mJ^\n_s-k^\n_s \mJ^\m_s+\frac{s}{2}\e^{\m\n\a\b}\nabla^s_{\a}\mJ^s_{\b}=0,
  \ee
where the chiral components $\mJ^\m_s$ of the Wigner function are defined as
\be
 \mJ^\m_s=\frac{1}{2}\(\mV^\m+s \mA^\m\),
  \ee
with $s=+$ and $s=-$ for right-handed and left-handed fermions respectively.
$k^s_{\m}=p_\m-A^s_\m$ is the kinetic momentum with the gauge potential in the chiral basis being $A^s_\m=A_\m+sA^5_\m$. $\nabla^s_\m=\pd_\m+(\pd_\n A^s_{\mu})\pd^\n_p$ is the covariant derivative at $O(\pd_X)$. Note that the kinetic momenta and covariant derivatives differ for the right and left handed components in the presence of axial gauge field.

\section{Solution to the Kinetic Equations}\label{sec_3}

We solve ${\mJ_\m}$ by gradient expansion up to first order,
\be
 \mJ^s_\m=\mJ^{s\0}_\m+\mJ^{s\1}_\m+\cdots ,
  \ee
where the superscripts $\0,\1,\cdots$ denotes the orders in the expansion. Substituting this expansion into Eqs.(\ref{b7})-(\ref{b9}) and requiring that the equations hold order by order, the equations for $\mJ^{s(n)}_\m$ with $n=1$ and $n=0$ read
\be\label{c1}
 k^s_\m \mJ^{s(n)\m}=0,
 \ee
\be\label{c2}
 \nabla^s_\m \mJ^{s(n)\m}=0,
  \ee
\be\label{c3}
 k_\m^s\mJ^{s(n)}_\n-k_\n^s \mJ^{s(n)}_\m+\frac{s}{2}\e_{\m\n\a\b}\nabla_s^{\a}\mJ^{s(n-1)\b}=0,
  \ee
where we have defined $\mJ^{s(-1)}_\m=0$. Contracting both sides of Eq.(\ref{c3}) with $k^\m_s$ and using Eq.(\ref{c1}), we have
\be\label{c4}
 k_s^2\mJ^{s(n)}_\m=\frac{s}{2}\e_{\m\n\a\b}k_s^\n \nabla_s^{\a}\mJ^{s(n)\b}.
  \ee
The solution of Eq.(\ref{c4}) implies a general form of $\mJ^{s(n)}_\m$:
\be\label{c5}
 \mJ^{s(n)}_\mu=J^{s(n)}_\m \delta (k_s^2)+\frac{s}{2k_s^2}\e_{\m\n\a\b}k_s^\n \nabla_s^{\a}\mJ^{s(n-1)\b},
  \ee
where $J^{s(n)}_\m$ is nonsingular at $k_s^2=0$, and can be constrained by substituting Eq.(\ref{c5}) back into Eq.(\ref{c3}). From Eq.(\ref{c1}), we can also obtain a further constraint for $J^{s(n)}_\m$: $k_s^\m J^{s(n)}_\m \delta(k_s^2)=0$.

When $n=0$, the equation of motion depend on kinetic momentum only. It is straightforward to write down the solution,
\be\label{c6}
 \mJ^{s\0}_{\m}=k^s_\m\delta(k_s^2)f_s(k_s,X).
  \ee
To be specific, in this paper we choose the zeroth order distribution function to be Fermi-Dirac distribution in kinetic momentum
\be\label{c7}
 f_s(k_s,X)=\frac{2}{(2\pi)^3}\left[\theta(u\cdot k_s)f_{FD}(\beta\cdot k_s -\bar\mu_s)+\theta(-u\cdot k_s)f_{FD}(-\beta\cdot k_s+\bar\mu_s)\right],
   \ee
where the Fermi-Dirac distribution function $f_{FD}(z)\equiv [exp(z)+1]^{-1}$, $\b^\mu\equiv\beta u^\m$ with $\b=\b^\m u_\m\equiv 1/T$ being the inverse temperature and $u^\mu$ being the fluid velocity, and $\bar\m_s=\beta \m_s$ with $\m_s=\m+s\m_5$ being the chemical potential for chirality $s=\pm 1$. The hydrodynamic quantities $\b^\m$, $\b$ and $\bar{\m}_s$ can be dependent on $X$.

Using the definitions of $k^s_\m$ and $\nabla^s_\m$ given in Sec.II, we can derive the following useful identities,
\be\label{c16}
 \begin{split}
  \nabla^s_\m k^s_\n&=-F^s_{\m\n},\\
   \nabla^s_\m \delta^{(n)}(k_s^2)&=-2 F^s_{\m\n}k_s^\n \delta^{(n+1)}(k_s^2),
    \end{split}
     \ee
where we have defined $F^s_{\m\n}\equiv\pd_\m A^s_\n-\pd_\n A^s_\m =F_{\m\n}+s F^5_{\mu\nu}$ and $\delta^{(n)} (k^2)=\(\frac{d}{d k^2}\)^{n}\delta(k^2)$. Especially, it follows that $\nabla^s_\m\left[k^\m_s \delta(k_s^2)\right]=0$. Using this identity, it is easy to show that
\be\label{c8}
 \begin{split}
  \nabla^s_\m \mJ_s^{\0\m}&=\nabla_\m\left[k_s^\m\delta(k_s^2)f_s(X,k_s)\right]\\
                       &=\delta(k_s^2)k_s^\m\nabla_\m f_s(X,k_s).
    \end{split}
     \ee
It implies the constraint equation $k_s^\m\nabla_{s,\m} f_s =0$ in order for Eq.(\ref{c2}) to hold. With the distribution function given by Eq.(\ref{c7}), we can evaluate $k_s^\m\nabla_{s,\m} f_s =0$ as
\be\label{c9}
 \begin{split}
  k_s^\m \nabla^s_{\m} f_s&=\frac{\pd f_s}{\pd (\b\cdot k_s)}k_s^\mu\left[\nabla_\m(\b\cdot k_s)-\nabla^s_\m \bar\m_s\right]\\
                   &=\frac{\pd f_s}{\pd (\b\cdot k_s)}k_s^\mu \left[k_s^\n \pd_\m \beta_\n -\pd_\m \bar\m_s- F^s_{\m\n} \beta^\n\right].
     \end{split}
      \ee
Accordingly we deduce the conditions for the chiral systems to be in global equilibrium, that is
\be\label{c10}
 \pd_\m \b_\n+\pd_\n\b_\m=0,
  \ee
\be\label{c11}
 \pd_\m\bar\m_s=-F^s_{\m\n}\b^\n.
  \ee
Eq.(\ref{c10}) is the Killing condition for $\b_\m$ which leads to the solution $\b_\m=b_\m-\Omega_{\m\n}x^\n$ with $b_\m$ and the thermal vorticity $\Omega_{\m\n}=\frac{1}{2}(\pd_\m \b_\n-\pd_\n \b_\m)$ being constants. Eq.(\ref{c11}) takes a more familiar form in the original basis. With $\bar\mu=\frac{1}{2}(\bar\m_+ +\bar\m_-)$ and $\bar\mu_5=\frac{1}{2}(\bar\m_+ -\bar\m_-)$, we have
\be\label{c27}
\pd_\m \bar\m=-F_{\m\n}\b^\n,
\ee
\be\label{c28}
\pd_\m \bar\m_5 =-F^5_{\m\n} \b^\n.
\ee
Eq.\eqref{c27} implies the electric field is balanced by gradient of chemical potential. Eq.(\ref{c28}) is just the axial counterpart of Eq.(\ref{c27}).

With the help of the conditions given by Eqs.(\ref{c10}) and (\ref{c11}), we can derive that
\be\label{c12}
 \nabla_\m f_s= f_s^\prime \Omega_{\m\n}k_s^\n,
  \ee
where $f_s^\prime\equiv\frac{\pd f_s}{\pd(\b\cdot k_s)}$. Substituting the zeroth order solution Eq.(\ref{c6}) into Eq.(\ref{c5}) gives the first order solution ($n=1$):
\be\label{c13}
 \begin{split}
  \mJ^{s\1}_\m&=J^{s\1}_\m \delta(k_s^2)+\frac{s}{2k_s^2}\e_{\m\n\a\b}k_s^\n \nabla_s^\a \mJ^{s\0\b}\\
            &=J^{s\1}_\m \delta(k_s^2)+s\tilde F^s_{\m\n}k_s^\n f_s\delta^\prime(k_s^2),
    \end{split}
     \ee
where we have used $\tilde F^s_{\m\n}=\frac{1}{2}\e_{\m\n\a\b}F_s^{\a\b}$ and $\delta^\prime(k_s^2)=-\delta(k_s^2)/k_s^2$. Then, substituting Eqs.(\ref{c6}) and (\ref{c13}) into Eq.(\ref{c3}), we arrive at
\be\label{c33}
 k^s_\m\( J^{s\1}_\n \delta(k_s^2)+s \tilde F^s_{\n\r}k_s^\r f_s\delta^\prime(k_s^2)\)-[\m \leftrightarrow \n]=-\frac{s}{2}\e_{\m\n\a\r}\nabla_s^\a\left[k_s^\r f_s \delta(k_s^2)\right].
  \ee
It help us to determine $J^{s\1}_\m=-\frac{s}{2}\tilde\Omega_{\m\n}k_s^\n f_s^\prime$. Details of the determination are presented in Appendix A.
We summarize the solution of the Wigner function up to the first order:
\be\label{c18}
 \mJ^{s}_\m=k^\m_s \delta(k_s^2)f_s-\frac{s}{2}\tilde\Omega_{\m\n}k_s^\n \delta(k_s^2)f_s^\prime+s\tilde F^s_{\m\n}k_s^\n \delta^\prime(k_s^2)f_s.
  \ee

Eq.\eqref{c18} generalizes the known solution to the case with axial gauge fields \cite{Yang:2020mtz}. In fact, we can show they are equivalent upon proper identification. In the absence of axial gauge field, the vector gauge link is inserted as
\begin{align}\label{link}
\overline{S}_{\text{link}}^<(x,y)=\overline{S}^<(x,y)e^{-i\int_y^x dz\cdot A(z)}=\overline{S}^<(x,y)e^{-is\cdot A(X)}+O(\pd_X^2).
\end{align}
We have used the subscript $\text{link}$ to indicate that the corresponding quantities have explicit gauge link insertion.
The vector gauge invariant $\overline{S}_{\text{link}}^<(x,y)$ is then Wigner transformed with kinetic momentum:
\begin{align}\label{link2}
{S}_{\text{link}}^<(k,X)&=\int d^4s e^{ik\cdot s}\overline{S}_{\text{link}}^<(x,y),
\end{align}
generating the gauge invariant observables from components of Wigner function. Importantly the momentum appearing in Eq.\eqref{link2} should be the gauge invariant kinetic momentum $k$. We choose to work with bare Wigner function, which is then transformed with canonical momentum $p$. From Eq.\eqref{link}, we can easily obtain
\begin{align}
e^{ik\cdot s}\overline{S}_{\text{link}}^<(x,y)=e^{ip\cdot s}\overline{S}^<(x,y)\no
\Rightarrow {S}_{\text{link}}^<(k,X)=S^<(p,X).
\end{align}
It shows that up to $O(\pd_X)$, our bare Wigner function is also vector gauge invariant: the gauge dependence in our bare Wigner function is canceled by the gauge dependence in the canonical momentum in the Wigner transform. Therefore components of our $S^<(p,X)$ is also gauge invariant, whose momentum integration give rise to physical observables.
When the axial gauge field is present, it is straightforward to deduce $S^<(p,X)$ is invariant under both vector and axial gauges.
In fact, we may equivalently working with Wigner functions for right and left-handed fermions with appropriate gauge link insertion:
\begin{align}
&\overline{S}_s(x,y)=\langle\ps_s(x)\ps_s^\dg(y)\rangle,\no
&\overline{S}_{s,\text{link}}(x,y)=S^i(x,y)e^{-i\int_y^x dz\cdot A_s(z)},
\end{align}
with $\ps_+\equiv\ps_R$ and $\ps_-\equiv\ps_L$ standing for right and left handed fermions respectively.

\section{Anomalous Transports and Correlation Functions}\label{sec_4}
Integrating Eq.(\ref{c18}) over the kinetic momentum $k_s$, we obtain the left-handed and right-handed currents $j^\m_s(X)$ up to the first order in $\pd_X$
\be\label{d1}
j_s^\m(X)=\int d^4k_s \mJ^\m_s(k_s ,X),
\ee
Both the solution Eq.\eqref{c18} and integration measure are vector/axial gauge invariant, it follows that the resulting currents are gauge invariant.

After the four-momentum integrations, we have the zeroth order contribution given by $\mJ_s^{\0 \m}$, and the first order contribution given by $\mJ^{\1\m}_s$:
\be\label{d2}
j^{\0\m}_s=n_s u^\mu,
\ee
\be\label{d3}
j^{\1\m}_s=\x_s \omega^\m +\x_{Bs}B_s^\mu,
\ee
with $n_s$ being the fermion number density, $\x_s$ and $\x_{Bs}$ related to transport coefficients of chiral vortical effect (CVE), chiral magnetic effect (CME) and chiral separation effect (CSE). The vorticity and magnetic field are defined as $\o^\mu=T\tilde \Omega^{\m\n}u_\n$ and $B_s^\m=\tilde F_s^{\m\n} u_\n$ respectively according to the decomposition,
\be\label{d4}
\tilde F^s_{\m\n}=B^s_\m u_\n- B^s_\n u_\m-\e_{\m\n\a\b}u^\a E^\b_s,\;\; F^s_{\m\n}=E^s_\m u_\n -E^s_\n u_\m+\e_{\m\n\a\b}u^\a B_s^{\b},
\ee
\be\label{d5}
T\tilde\Omega_{\m\n}=\o_\m u_\n-\o_\n u_\m-\e_{\m\n\a\b}u^\a \varepsilon^\b,\;\; T\Omega_{\m\n}=\varepsilon_\m u_\n-\varepsilon_\n u_\m+\e_{\m\n\a\b}u^a \o^\b.
\ee
And the coefficients $n_s$, $\x_s$ and $\x_{Bs}$ are given by
\be\label{d6}
n_s =\frac{\mu_s}{6 \pi^2}\(\frac{\pi^2}{\b^2}+\m_s^2\),
\ee
\be\label{d7}
\x_s=\frac{s}{12\pi^2}\(\frac{\pi^2}{\b^2}+3\m_s^2\).
\ee
\be\label{d8}
\x_{Bs}=\frac{s}{4\pi^2}\m_s
\ee
The vector and axial currents can be obtained from the linear combinations of $j^\m_\pm$:
\be
 j^\m=j^\m_++j^\m_-,\;\;\;\;\;j^\m_5=j^\m_+-j^\m_-.
  \ee
Then the zeroth order vector and axial currents read
\be
 j^{\0\m}=n u^\m,
  \ee
   \be
    j_5^{\0\m}=n_5 u^\m.
      \ee
From the first order currents, we can obtain the vector and axial currents in the CVE
\be\label{d9}
j^\m_\o=\x \o^\m,
\ee
\be\label{d10}
j^\m_{5,\o}=\x_5 \o^\m,
\ee
while the vector and axial currents in the CME are
\be\label{d11}
j^\m_B=\x_B B^\m+\x_{B5} B^\m_5,
\ee
\be\label{d12}
j^\m_{5,B}=\x_{B5} B^\m+\x_B B_5^\m,
\ee
where
\be\label{d13}
\begin{split}
&n=\frac{\m}{3\pi^2}\(\frac{\pi^2}{\b^2}+\m^2+3\m_5^2\), \;\;\;\;n_5=\frac{\m_5}{3\pi^2}\(\frac{\pi^2}{\b^2}+3\m^2+\m_5^2\),\\
&\x=\frac{\m\m_5}{\pi^2},\;\;\x_5=\frac{1}{6\b^2}+\frac{\m^2}{2\pi^2}+\frac{\m^2_5}{2\pi^2},\;\;\x_B=\frac{\m_5}{2\pi^2},\;\;\x_{B5}=\frac{\m}{2\pi^2}.
\end{split}
\ee

The gauge invariant canonical energy-momentum tensor can be obtained by
\be\label{d14}
 \begin{split}
   T^{\m\n}&=\int d^4k_+ \mJ^\m_+(k_+,X) k_+^\n+\int d^4k_- \mJ^\m_-(k_-,X) k_-^\n\\
   &\equiv T_+^{\m\n}+T_-^{\m\n}.
   \end{split}
  \ee
Then we can perform the four momentum integrals to obtain
\be\label{d15}
 T^{\0\m\n}_s =\(u^\m u^\n  -\frac{1}{3}\Delta^{\m\n}\)\rho_s,
  \ee
\be\label{d16}
 T^{\1\m\n}_s = s n_s (u^\m \o^\n+u^\n \o^\m)+\frac{\x_s}{2}(u^\m B_s^\n+u^\n B_s^\m-\e^{\m\n\a\b}u_\a E^s_\b)-\frac{s n_s}{2}(u^\m \o^\n-u^\n \o^\m+\e^{\m\n\a\b}u_\a \varepsilon_\b),
   \ee
with
\be\label{d17}
 \r_s=\frac{7}{120}\frac{\pi^2}{\b^4}+\frac{1}{4}\frac{\m_s^2}{\b^2}+\frac{1}{8\pi^2}\m_s^4.
  \ee
Eqs.\eqref{d16} and \eqref{d17} generalize the first order results in \cite{Yang:2020mtz} to the case with axial gauge field.
We can further separate the symmetric and anti-symmetric parts of the canonical energy-momentum tensor as\footnote{Implicitly we assume that the fluid velocity is not redefined in the presence of external sources including vector/axial gauge fields and vorticity. This chooses a particular hydrodynamic frame, termed thermodynamic frame in literature \cite{Jensen:2012jh}.}
\be\label{d18}
 T^{\0\{\m\n\}}=\(u^\m u^\n  -\frac{1}{3}\Delta^{\m\n}\)\rho,
  \ee
\begin{align}\label{d19}
  T^{\1\{\m\n\}}&=n_5 (u^\m \o^\n+u^\n \o^\m)
      +\frac{\x}{2}(u^\m B^\n+u^\n B^\m)+\frac{\x_5}{2}(u^\m B_5^\n+u^\n B_5^\m),
      \end{align}
\begin{align}\label{d22}
  T^{\1[\m\n]}&=-\frac{\x}{2}\e^{\m\n\a\b}u_\a E_\b-\frac{\x_5}{2}\e^{\m\n\a\b}u_\a E^5_\b-\frac{n_5}{2}(u^\m \o^\n-u^\n \o^\m+\e^{\m\n\a\b}u_\a \varepsilon_\b),
     \end{align}
where $T^{\{\mu\nu\}}\equiv \frac{T^{\m\n}+T^{\n\m}}{2}$ and $T^{[\mu\nu]}\equiv \frac{T^{\m\n}-T^{\n\m}}{2}$. $\r=\frac{7}{60}\frac{\pi^2}{\beta^4}+\frac{1}{2}\frac{\m^2+\m_5^2}{\beta^2}+\frac{1}{4\pi^2} \(\m^4+6\m^2\m^2_5+\m_5^4\)$ is the energy density. The RHS of Eq.\eqref{d19} corresponds to heat flow along vorticity, magnetic and axial magnetic fields respectively. The first two need chiral imbalance in the fluid to exist while the last one exists even in the neutral medium. As we shall see shortly, the last one can be related to axial chiral vortical effect by Onsager relation. Using the Killing conditions, we show in Appendix B the following conservation equations
\be\label{d20}
\pd_\m j^{\m}=\frac{1}{2\pi^2}\(\vec E_5\cdot \vec B+\vec E\cdot \vec B_5\),
\ee
\be\label{d21}
\pd_\m j^{\m}_{5}=\frac{1}{2\pi^2}\(\vec E\cdot \vec B+\vec E_5\cdot \vec B_5\),
\ee
\be
\begin{split}\label{Fj}
\pd_\m T^{\{\m\n\}}=F^{\n\m}j_\m+F_5^{\m\n}j_{5,\mu},
  \end{split}
\ee
\be\label{Tcons}
\pd_\m T^{[\m\n]}=0,
\ee
\be\label{spin_tensor}
T^{[\m\n]}=\pd_\l S^{\l\m\n}.
\ee
Eqs.\eqref{d20} and \eqref{d21} are current conservation equations. Eqs.\eqref{Fj} and \eqref{Tcons} are the energy-momentum conservation subject to external force by vector and axial gauge fields. Eq.\eqref{spin_tensor} corresponds to change rate of spin tensor $S^{\l\m\n}$.

We will be mainly interested in correlation functions among vector/axial currents, which are obtainable by functional derivatives with respect to vector/axial gauge potential. For this purpose, we may turn off the vorticity and set $u^\m=(1,0,0,0)$. The Killing condition $\pd_\m\b_\n+\pd_\n\b_\m=0$ implies a homogeneous temperature.

It is known that the definition of current is not unique when axial gauge field is present. One can choose either consistent current and covariant current \cite{Landsteiner:2016led}, with the former always conserve vector current and the latter is symmetric with respect to the interchange of vector/axial components. The construction of our solution suggests the corresponding current to be covariant current. Indeed, the anomaly equations we obtained Eqs.\eqref{d20} and \eqref{d21} agree with those of covariant current.

Now we turn to the calculation of correlation functions. A convenient way to calculate correlation function is to take functional derivatives with respect to vector/axial gauge potential. Note that each functional derivative brings down a consistent current as
\begin{align}
\frac{\d }{\d A_\m(x)}e^{i\G[A,A_5]}=j^\m_{\text{cons}}(x)e^{i\G[A,A_5]},\quad \frac{\d }{\d A^5_\m(x)}e^{i\G[A,A5]}=j^\m_{5,\text{cons}}(x)e^{i\G[A,A5]},
\end{align}
with $\G[A,A_5]$ being the effective action. Taking multiple derivatives give multi-point correlation functions. We illustrate this with examples of two and three point functions.
Instead of finding the effective action, we start with one-point function of consistent currents, which are related to the covariant currents in Eqs.\eqref{d11} and \eqref{d12} by \cite{Landsteiner:2016led}
\begin{align}\label{cons_cov}
j_{\text{cov}}^\m&=j_{\text{cons}}^\m+\frac{1}{4\p^2}\e^{\m\n\r\s}A_{5,\n}F_{\r\s},\no
j_{5,\text{cov}}^\m&=j_{5,\text{cons}}^\m+\frac{1}{12\p^2}\e^{\m\n\r\s}A_{5,\n}F_{5\r\s}.
\end{align}
From Eqs.\eqref{d11} and \eqref{d12}, we obtain
\begin{align}\label{jcons}
j_{\text{cons}}^i&=\frac{1}{2\p^2}(\m_5-A_0^5)B_i+\frac{1}{2\p^2}\m B_5^i,\no
j_{5\text{cons}}^i&=\frac{1}{2\p^2}(\m_5-\frac{A_0^5}{3})B^5_i+\frac{1}{2\p^2}\m B^i.
\end{align}
Possible contribution from vector/axial electric fields are not included in Eq.\eqref{jcons} for the following reason: they need to be balanced by gradients of corresponding chemical potentials. Since we will calculate correlation functions in system with constant chemical potential and temperature, we simply turn them off.
We will calculate correlation functions for equilibrium state without axial gauge field, i.e. $A_{5,\m}=0$. Eq.\eqref{jcons} indicates the only nonvanishing correlation functions are two-point and three-point ones. The former comes from CME and CSE terms (and their analogs with axial magnetic field). Fourier transforms of these terms give
\begin{align}
\tilde{j}_{\text{cons}}^i(k)&=\frac{1}{2\p^2}\m_5\tilde{B}_i(k)+\frac{1}{2\p^2}\m \tilde{B}_5^i(k),\no
\tilde{j}_{5,\text{cons}}^i(k)&=\frac{1}{2\p^2}\m_5\tilde{B}^5_i(k)+\frac{1}{2\p^2}\m \tilde{B}^i(k),
\end{align}
where we use tilde to indicate operators in Fourier space.
The absence of electric fields requires momenta appearing in the Fourier transforms contain no temporal components.
Taking functional derivative once, and noting the coupling $\int d^4x A_\m(x)j^\m_{\text{cons}}(x)=\int\frac{d^4k}{(2\p)^4}\tilde{A}_\m(k)\tilde{j}_{\text{cons}}^\m(-k)$ and similarly for axial counterpart, we obtain
\begin{align}\label{2point}
\langle\tilde{j}_{\text{cons}}^i(k)\tilde{j}_{\text{cons}}^j(-k)\rangle&=\frac{i\m_5\e^{ijk}k^k}{2\p^2},\no
\langle\tilde{j}_{\text{cons}}^i(k)\tilde{j}_{5,\text{cons}}^j(-k)\rangle&=\frac{i\m\e^{ijk}k^k}{2\p^2},\no
\langle\tilde{j}_{5,\text{cons}}^i(k)\tilde{j}_{5,\text{cons}}^j(-k)\rangle&=\frac{i\m_5\e^{ijk}k^k}{2\p^2},\no
\langle\tilde{j}_{5,\text{cons}}^i(k)\tilde{j}_{\text{cons}}^j(-k)\rangle&=\frac{i\m\e^{ijk}k^k}{2\p^2},
\end{align}
where the LHS are defined by $\langle X(k)Y(p)\rangle=\langle X(k)Y(-k)\rangle(2\p)^4\d^{(4)}(k+p)$.
The terms quadratic in $A(A_5)$ give rise to the three-point correlation functions. Fourier transforms of these terms are given by
\begin{align}
\tilde{j}_{\text{cons}}^i(q)&=\int\frac{d^4k}{(2\p)^4}\big[-\frac{1}{2\p^2}\tilde{A}_0^5(q-k)\tilde{B}_i(k)\big],\no
\tilde{j}_{5,\text{cons}}^i(q)&=\int\frac{d^4k}{(2\p)^4}\big[-\frac{1}{6\p^2}\tilde{A}_0^5(q-k)\tilde{B}_i^5(k)\big].
\end{align}
Taking functional derivatives twice, we obtain the following three-point correlation functions
\begin{align}\label{3point}
\langle \tilde{j}_{cons}^i(q)\tilde{j}_{5,cons}^0(k-q)\tilde{j}_{cons}^j(-k)\rangle&=-i\e^{ijk}\frac{k^k}{2\p^2},\no
\langle \tilde{j}_{5,cons}^i(q)\tilde{j}_{5,cons}^0(k-q)\tilde{j}_{5,cons}^j(-k)\rangle&=-i\e^{ijk}\frac{k^k}{6\p^2},
\end{align}
where the LHS are defined by $\langle X(k)Y(q)Z(p)\rangle=\langle X(k)Y(q)Z(-k-q)\rangle(2\p)^4\d^{(4)}(k+q+p)$.
In the limit $q\to k$, Eq.\eqref{3point} are in agreement with the results obtained with field theory and holography \cite{Landsteiner:2011cp}.

We can also calculate correlation functions between energy-momentum tensor and currents. From Eq.\eqref{d19} and the fact $E_\m=E_{5,\m}=0$, it is clear the nonvanishing correlation functions involves only symmetric part of the energy-momentum tensor:
\begin{align}\label{TJ}
\langle T^{\{0i\}}(k)j_{cons}^j(-k)\rangle&=\frac{\x}{2}\e^{ijk}ik^k,\no
\langle T^{\{0i\}}(k)j_{5,cons}^j(-k)\rangle&=\frac{\x_5}{2}\e^{ijk}ik^k,
\end{align}
Indeed, they can be related to CVE in Eq.\eqref{d10}. Noting that $\o$ can be induced by metric perturbation as
\begin{align}
\o^j=-\frac{1}{2}\e^{ijk}\pd_kh_{0i},
\end{align}
we find from Eqs.\eqref{d9} and \eqref{d10}
\begin{align}\label{JT}
\langle j_{cons}^j(k)T^{\{0i\}}(-k)\rangle&=\frac{\x}{2}\e^{ijk}ik^k,\no
\langle j_{5,cons}^j(k)T^{\{0i\}}(-k)\rangle&=\frac{\x_5}{2}\e^{ijk}ik^k.
\end{align}
Eq.\eqref{JT} and Eq.\eqref{TJ} are consistent with Onsager relation\footnote{See \cite{Bu:2019qmd} for the case with a background magnetic field.}.



\section{Summary and Outlook}\label{sec_5}

We have derived chiral kinetic theory with both vector and axial gauge fields. We have also found a solution preserving both vector and axial gauge symmetry, leading to covariant vector and axial currents. The introduction of axial gauge field allows us to derive correlation functions of vector/axial current. This is done by first converting covariant currents to consistent current and then taking functional derivatives with respect to vector/axial gauge fields. We find the resulting correlation functions in agreement with known field theoretic results.

The present work can be extended in two aspects. Firstly, mass effect can be included. Since mass breaks axial symmetry explicitly, it requires non-trivial modification to the solution presented in this work. It is known that mass introduces additional degrees of freedoms. It would be interesting to see how the dynamics of these degrees of freedoms are affected by the presence of axial gauge field. More interestingly, collisional effect should be included for a complete description of dynamics, in particular for axial current (or spin density). This would provide a route to spin hydrodynamics for weakly interacting fermion system.

\appendix

\section{Verification of Eqs.\eqref{c2} and determination of $J^{s\1}_\m$}
We determine $J^{s\1}_\m$ from Eq.\eqref{c33}:
\be\label{c14}
 \begin{split}
  \(k^s_\m J^{s\1}_\n-k^s_\n J^{s\1}_\m\)\delta(k_s^2)
  &=-s\(k^s_\m \tilde F^s_{\n\r}-k^s_\n \tilde F^s_{\m\r}\)k_s^\r f_s\delta^\prime(k_s^2)-\frac{s}{2}\e_{\m\n\a\r} \nabla_s^\a \left[k_s^\r f_s \delta(k_s^2)\right]\\
                                            &=-s\(k^s_\m \tilde F^s_{\n\r}-k^s_\n \tilde F^s_{\m\r}+k^s_\r \tilde F^s_{\m\n}\)k_s^\r f_s\delta^\prime(k_s^2)+\frac{s}{2}\e_{\m\n\r\a}k_s^\r \nabla_s^\a \left[f_s \delta(k_s^2)\right]\\
                                            &=s \e_{\m\n\r\a}F^{\a\b}_s k^s_\b k_s^\r f_s \delta^\prime (k_s^2)+\frac{s}{2}\e_{\m\n\r\a}k_s^\r \nabla_s^\a \left[f_s \delta(k_s^2)\right]\\
                                            &=\frac{s}{2}\e_{\m\n\r\a}k_s^\r \Omega^{\a\b}k^s_\b f_s^\prime \delta(k_s^2),
         \end{split}
          \ee
where we have used Eqs.(\ref{c16},\ref{c12}), and the Schouten identity
\be\label{c15}
 g^{\r}_\s\e^{\m\n\a\b}+g^{\b}_\s\e^{\r\m\n\a}+g^{\a}_\s\e^{\b\r\m\n}+g^{\n}_\s \e^{\a\b\r\m} +g^{\m}_\s \e^{\n\a\b\r}=0,
  \ee
from which we can show that $k_s^\m \tilde F_s^{\n\r}+k_s^\r \tilde F_s^{\m\n}+k_s^\n \tilde F_s^{\r\m}+\e^{\m\n\r\a} F^s_{\a\b}k_s^\b=0$ by contracting both sides with $F^s_{\a\b}k_\s$ and using the antisymmetric nature of the field strength tensors. At this stage, we can express the tensor in terms of the dual tensor: $\Omega^{\a\b}=-\frac{1}{2}\e^{\a\b\k\l}\tilde\Omega_{\k\l}$. And the Levi-Civita tensors can be contracted as
\be
 \begin{split}\label{c20}
  \e^{\m\n\r\s}\e_{\m\n\a\b}&=-2! \;\delta^{\r\s}_{\a\b}=-2(\d^\r_\a \d^\s_\b-\d^\r_\b\d^\s_\a),\\
   \e^{\m\k\r\s}\e_{\m\n\a\b}&=-\d^{\k\r\s}_{\n\a\b}=-\d^\k_\n( \d^\r_\a \d^\s_\b- \d^\r_\b \d^\s_\a)
                            -\d^\k_\a(\d^\r_\b\d^\s_\n-\d^\r_\n\d^\s_\b)-\d^\k_\b(\d^\r_\n\d^\s_\a-\d^\r_\a\d^\s_\n) .
      \end{split}
       \ee
Using these contract relations, we can further derive for two arbitrary antisymmetric tensors, e.g. $\Omega_{\m\n}$ and $F^s_{\m\n}$, that
\be\label{c21}
 \frac{1}{2}\tilde\Omega^{\m\n}F^s_{\m\n}k_s^2=\Omega^{\m\a}\tilde F^s_{\m\r}k^s_\a k_s^\r +\tilde\Omega^{\m\a} F^s_{\m\r}k^s_\a k_s^\r,
  \ee
and especially,
\be\label{c22}
 \begin{split}
  \tilde\Omega^{\m\n}\Omega_{\m\n}k_s^2&=4\Omega^{\m\a}\tilde \Omega_{\m\r}k^s_\a k_s^\r, \\
   \tilde F_s^{\m\n}F^s_{\m\n}k_s^2&=4F_s^{\m\a}\tilde F^s_{\m\r}k^s_\a k_s^\r.
    \end{split}
     \ee
Then Eq.(\ref{c14}) can be written as
\be\label{c17}
 \begin{split}
  \(k^s_\m J^\1_\n-k^s_\n J^\1_\m\)\delta(k_s^2)&=\frac{s}{2}\e_{\m\n\r\a}k_s^\r \Omega^{\a\b}k^s_\b f_S^\prime \delta(k_s^2)\\
  &=-\frac{s}{4}\e_{\m\n\r\a}\e^{\a\b\k\l} \tilde\Omega_{\k\l}k_s^\r k^s_\b f_s^\prime \delta(k_s^2)\\
                                          &=-\frac{s}{4}\delta^{\b\k\l}_{\m\n\r}k_s^\r \tilde\Omega_{\k\l} k^s_\b f_s^\prime \delta(k_s^2)\\
                                          &=-\frac{s}{2}\(k^s_\m \tilde\Omega_{\n\r}k_s^\r-k^s_\n \tilde\Omega_{\m\r}k_s^\r\)f^\prime\delta(k_s^2).
    \end{split}
     \ee
We have thrown away the terms proportional to $k_s^2 \delta(k_s^2)=0$. In fact, Eq.(\ref{c17}) cannot exclude the correction of the form $\delta J^\1_\m= k^s_\m X^\1$, which involves an unknown regular function $X^\1$ that should be identified as of the first order in gradient. We may absorb this undetermined correction into the first order distribution function which we do not consider in this work. After that, we can set $J^{s\1}_\m=-\frac{s}{2}\tilde\Omega_{\m\n}k_s^\n f_s^\prime$ in line with Eq.(\ref{c17}).

Now we check if the first-order solution given by Eq.(\ref{c18}) satisfies Eq.(\ref{c2}) with the global equilibrium conditions which have been embedded into Eq.(\ref{c12}). For convenience, we divide the Wigner function into two parts, $\mathcal{J}^{s\1\m}_\O$ and $\mathcal{J}^{s\1\m}_{EM}$, which are generated by vorticity and background fields respectively. We note that the vorticity $\Omega_{\m\n}$ is constant due to the Killing condition $\pd_\m \b_\n+\pd_\n \b_\m=0$. Assuming the constant field strength tensors, i.e. $\pd_\r F^s_{\m\n}=0$, we can show that
\be\label{c23}
 \begin{split}
  \nabla^s_\m \mJ_\Omega^{s\1\m}&=-\frac{s}{2}\tilde\Omega^{\m\n}\nabla^s_\m\left[k^s_\n f_s^\prime \delta(k_s^2)\right]\\
                             &=\frac{s}{2}\tilde\Omega^{\m\n}F^s_{\m\n}f_s^\prime \delta(k_s^2) +s\tilde\Omega^{\m\n}F^s_{\m \r}k^s_\n k_s^\r f_s^\prime \d^\prime(k_s^2)
                             -\frac{s}{2}\tilde\Omega^{\m\n} \Omega_{\m\r}k^s_\n k_s^\r f_s^{\prime\prime} \d(k_s^2)\\
                             &=-s\Omega^{\m\n}\tilde F^s_{\m \r}k^s_\n k_s^\r f_s^\prime \d^\prime(k_s^2),
     \end{split}
      \ee
      and
      \be\label{c24}
 \begin{split}
  \nabla^s_\m \mJ_{EM}^{s\1\m}&=s\tilde F_s^{\m\n}\nabla^s_\m\left[k^s_\n f_s\delta^\prime(k_s^2)\right]\\
                           &=-s\tilde F_s^{\m\n}F^s_{\m\n} f_s\delta^\prime(k_s^2)-2s\tilde F_s^{\m\n}F^s_{\m\r}k^s_\n k_s^\r f_s \d^{\prime\prime}(k_s^2)+s\tilde F_s^{\m\n}\Omega_{\m\r}k^s_\n k_s^\r f_s^\prime\delta^\prime(k_s^2)\\
                           &=s\tilde F_s^{\m\n}\Omega_{\m\r}k^s_\n k_s^\r f_s^\prime\delta^\prime(k_s^2).
   \end{split}
    \ee
where we have used $\delta^\prime(k^2)=-\d(k^2)/k^2$, $\d^{\prime\prime}(k^2)=-2 \d^\prime(k^2)/k^2$ and Eq.(A5).
By taking the sum of Eq.(A7) and Eq.(A8), we obtain
\be\label{c25}
 \nabla_\m \mJ^{\1\m}=0.
  \ee

\section{Conservation equations}
Using Eqs.(\ref{c10}) and (\ref{d5}) with the definition $\beta_\m\equiv\beta u_\m=u_\m/T$, we can easily show that
\be
\pd_\m \beta^\m\equiv\pd_\m \frac{u^\mu}{T}=0,
\ee
\be
\pd_\m \beta =u^\n \pd_\m \beta_\n-\beta u^\n\pd_\m u_\n=u^\n \Omega_{\m\n}=\beta \varepsilon_\m.
\ee
\be
u\cdot\pd \beta=0,\;\;\;\;\pd_\m u^\m=0
\ee
\be
\pd_\m u_\n=T\O_{\m\n}- \varepsilon_\m u_\n
\ee
And we have used $u\cdot \varepsilon=0$, $u^2=1$ and $u^\n\pd_\m u_\n=0$.
Then from Eqs.(\ref{c27}) and (\ref{c28}), we have
\be
\pd_\m \m=-\m \varepsilon_\m -F_{\m\n}u^\n=-\m \varepsilon_\m -E_\m,
\ee
\be
\pd_\m \m_5=-\m_5 \varepsilon_\m -F^5_{\m\n}u^\n=-\m_5 \varepsilon_\m -E^5_\m,
\ee
and the integrability conditions for constant $F_{\m\n}(F^5_{\m\n})$:
\be
F_{(5)\l}^{\;\m}\O^{\n\l}-F_{(5)\l}^{\;\n}\O^{\m\l}=0,
\ee
or equivalently,
\be
\e_{\m\n\a\b}(E_{(5)}^\a \o^\b-\varepsilon^\a B_{(5)}^\b)=0,\;\;\e_{\m\n\a\b}(E_{(5)}^\a \varepsilon^\b+\o^\a B_{(5)}^\b)=0.
\ee
Especially, we see that
\be
u\cdot \pd \m =u\cdot \pd \m_5=0.
\ee

Finally we can derive the following useful identities:
\be
\pd_\m n=-3 n \varepsilon_\m -2\xi_5 E_\m-2\xi E^5_\m,
\ee
\be
\pd_\m n_5= -3 n_5 \varepsilon_\m -2 \xi_5 E^5_\m-2\xi E_\m,
\ee
\be
\pd_\m \xi_5=-2 \xi_5 \varepsilon_\m -2 \xi_{B5} E_\m-2\xi_B E^5_\m,
\ee
\be
\pd_\m \xi=-2 \xi \varepsilon_\m-2 \xi_B E_\m-2\xi_{B5} E_\m^5,
\ee
\be
      \pd_\m \r= -4 \r \varepsilon_\m - 3 n E_\m -3 n_{5} E^5_\m,
\ee
\be
\pd_\m \xi_B =-\xi_B \varepsilon_\m-\frac{1}{2\pi^2}E^5_\m,\;\;\;\pd_\m \xi_{B5} =-\xi_{B5} \varepsilon_\m-\frac{1}{2\pi^2}E_\m,
\ee
which clearly give
\be
 u\cdot \pd \{n,n_5,\xi,\xi_5,\xi_B,\xi_{B5},\r\}=0.
 \ee

On the other hand, we can also show
\be
 \pd_\m \omega_\n=\varepsilon\cdot\omega g_{\m\n}-2 \varepsilon_\m \omega_\n,
  \ee
\be
 \pd_\m \varepsilon_\n=\o_\m\omega_\n-\varepsilon_\m\varepsilon_\n+\varepsilon^2 u_\m u_\n-\omega^2 \Delta_{\m\n}+\(u_\m\e_{\n\lambda\rho\s}+u_\n \e_{\m\l\r\s}\)u^\l \varepsilon^\r \o^\s,
   \ee
\be
 \pd_\m B_\n=-E_\m\omega_\n+\varepsilon\cdot B u_\m u_\n+\omega\cdot E \Delta_{\m\n}-\(u_\m\e_{\n\lambda\rho\s}+u_\n \e_{\m\l\r\s}\)u^\l \varepsilon^\r E^\s,
  \ee
\be
 \pd_\m E_\n=B_\m\omega_\n+\varepsilon\cdot E u_\m u_\n-\omega\cdot B \Delta_{\m\n}+\(u_\m\e_{\n\lambda\rho\s}+u_\n \e_{\m\l\r\s}\)u^\l E^\r \o^\s,
  \ee
where $\Delta^{\m\n}=g^{\m\n}-u^\m u^\n$. For example,
\be
\begin{split}
\pd_\m \o_\n &=\pd_\m\(T\tilde\Omega_{\n\r}u^\r\)=\tilde\Omega_{\n\r}\(u^\r\pd_\m T+T\pd_\m u^\r\)=\varepsilon\cdot\o g_{\m\n}-2\varepsilon_\m \o_\n.
\end{split}
\ee
These identities only hold with the constant $\Omega_{\m\n}$ and $F^s_{\m\n}$. Using Eqs.(B10)-(B20), we can verify the following conservation laws,
\be
 \pd_\m j^{(0)\m}=0,\;\;\;\;\pd_\m j^{(0)\m}_5=0,
  \ee
\be
\begin{split}
 \pd_\m j^{(1)\m}&= -\frac{1}{2\pi^2}\(E\cdot B_5+E_5\cdot B\)=\frac{1}{2\pi^2}\(\vec{E}\cdot\vec{B}_5+\vec{E}_5\cdot \vec{B}\),
\end{split}
\ee
\be
\begin{split}
 \pd_\m j^{(1)\m}_5&= -\frac{1}{2\pi^2}\(E\cdot B+E_5\cdot B_5\)=\frac{1}{2\pi^2}\(\vec{E}\cdot\vec{B}+\vec{E}_5\cdot \vec{B}_5\),\\
\end{split}
\ee
\be
\begin{split}
\pd_\m T^{\{\m\n\}}=F^{\n\m}j_\m+F_5^{\m\n}j_{5,\mu},
  \end{split}
\ee
\be
\pd_\m T^{[\m\n]}=0,
\ee
As an example, we evaluate the conservation law of the first-order symmetric energy current $T^{\1\{\m\n\}}$ given by Eq.(\ref{d19}). For convenience, we divide the energy current into two parts, $T^{\1\{\m\n\}}_\o$ and $T^{\1\{\m\n\}}_B$, which are generated by vorticity and background fields respectively. The derivative of $T^{\1\{\m\n\}}_\o$ can be evaluated as
\be
\begin{split}
\pd_\m T^{\1\{\m\n\}}_{\o}=&\pd_\m\left[n_5\(u^\m \o^\n+u^\n \o^\m\)\right]\\
    =&-2 \xi_5 \o\cdot E^5 u^\n-2 \xi \o\cdot E u^\n,
\end{split}
\ee
while the derivative of $T^{\1\{\m\n\}}_B$ is
\be
\begin{split}
\pd_\m T^{\1\{\m\n\}}_{B}=&\pd_\m\left[\frac{\xi}{2}\(u^\m B^\n+u^\n B^\m\)+\frac{\xi_5}{2}\(u^\m B_5^\n+u^\n B_5^\m\)\right]\\
    =&\xi \o\cdot E u^\n+\xi_5 \o\cdot E_5 u^\n+\xi\e_{\n\m\r\s} \o^\m u^\r B^\s +\xi_5\e_{\n\m\r\s} \o^\m u^\r B_5^\s \\
   &-u^\n E\cdot (\xi_B B+\xi_{B5}B_5)-u^\n E_5\cdot (\xi_{B5} B+\xi_{B}B_5),
\end{split}
\ee
Then we can verify the conservation law of the first-order symmetric energy-momentum tensor:
\be
\begin{split}
\pd_\m T^{\1\{\m\n\}}=&\pd_\m T^{\1\{\m\n\}}_{\o}+\pd_\m T^{\1\{\m\n\}}_{B}\\
     =&-u^\n E\cdot j-u^\n E_5\cdot j_5+\xi\e_{\n\m\r\s} \o^\m u^\r B^\s +\xi_5\e_{\n\m\r\s} \o^\m u^\r B_5^\s\\
     =&F^{\n\m}j_\m+F^{\n\m}_5 j_{5,\m},
\end{split}
\ee
where we have used the decomposition of Eq.(\ref{d4}) in the third line.

In addition to the conservation of currents and energy-momentum tensor, we can also obtain a relation between $T^{\1[\m\n]}$ and spin tensor $S^{\l\m\n}$. The latter is determined by the axial current as
\begin{align}
S^{\l\m\n}=\frac{1}{2}\e^{\eta\l\m\n}j_{5\eta}.
\end{align}
Integrating Eq.\eqref{b9} over $k_s$, we identify the first two terms as $T_s^{\1[\m\n]}$. The last term gives $\frac{s}{2}\e^{\m\n\a\b}\pd_\a j_\b^s$, while the term dependent on $A^s_\m$ inside $\nabla_\m^s$ simply drops out as a total derivative term. By taking the sum of right and left handed contributions, we arrive at
\begin{align}
T^{\1[\m\n]}=\pd_\l S^{\l\m\n}.
\end{align}

\begin{acknowledgments}
This work is in part supported by NSFC under Grant Nos 12075328, 11735007 and 11675274.
\end{acknowledgments}

\bibliographystyle{unsrt}
\bibliography{axial}

\end{document}